\documentclass[superscriptaddress,secnumarabic,amssymb,amsmath,nobibnotes,aps,prd,showkeys,showpacs,nofootinbib,preprint]{revtex4}

\usepackage{graphicx}
\usepackage{float}
\usepackage{bm}
\usepackage{amsmath}
\usepackage{amsfonts}
\usepackage{amssymb}
\usepackage{epstopdf}
\usepackage{natbib}%
\newcommand{\bee}{\begin{equation}}
\newcommand{\eee}{\end{equation}}
\newcommand{\eaa}{\end{eqnarray}}
\newcommand{\baa}{\begin{eqnarray}}

\def\ni{\noindent}

\begin{document}

\title{\Large Barrow black hole variable parameter model \\ connected to information theory}

\author{Everton M. C. Abreu}\email{evertonabreu@ufrrj.br}
\affiliation{Departamento de F\'{i}sica, Universidade Federal Rural do Rio de Janeiro, 23890-971, Serop\'edica, RJ, Brazil}
\affiliation{Departamento de F\'{i}sica, Universidade Federal de Juiz de Fora, 36036-330, Juiz de Fora, MG, Brazil}
\affiliation{Programa de P\'os-Gradua\c{c}\~ao Interdisciplinar em F\'isica Aplicada, Instituto de F\'{i}sica, Universidade Federal do Rio de Janeiro, 21941-972, Rio de Janeiro, RJ, Brazil}

%%%%%%%%%%%%%%%%%%%%%%%%%%%%%%%%%%%%%%%%%%%%%%%%%%%%%%%%%%%%%%%%%%%%%%%%%%%%%%%%%%%%%%%%%%%%

%%%%%%%%%%%%%%%%%%%%%%%%%%%%%%%%%%%%%%%%%%%%%%%%%%%%%%%%%%%%%%%%%%%%%%%%%%%%%%%%%%%%%%%%%%
\begin{abstract}
\ni One of the greatest challenges of theoretical physics today is to unveil the quantum information theory concerning what happens when one bit of information enters the black hole (BH) horizon.  The Landauer principle showed that a certain amount of energy is generated when one-bit of information is erased as it enters the event horizon system. In this paper we used the recently developed Barrow BH model to calculate the addition to the area of the event horizon of his toy model by using the Landauer concept. Besides we make this computation considering $\Delta$ as a constant and a variable parameter. We formulate the Barrow parameter ($\Delta$) as a function of the energy/mass, which is new in the Barrow BH literature.   We will investigate the differences between the Bekenstein-Hawking entropy ($\Delta=0$)  and the fractal ($\Delta=1$) cases concerning the addition in the area of the BH.  The asymptotical analysis is also mentioned and we will see that it affects only the fractal case.   All the results accomplished here are new concerning BHs in general and the Barrow model literature in particular.
\end{abstract}
%%%%%%%%%%%%%%%%%%%%%%%%%%%%%%%%%%%%%%%%%%%%%%%%%%%%%%%%%%%%%%%%%%%%%%%%%%%%%%%%%%%%%%%%%%%%
\date{\today}
\pacs{03.67.-a, 04.70.Dy, 04.70.-s}
\keywords{Black holes information theory, Landauer principle, Barrow black hole entropy}

\maketitle

\section{Introduction}

Entropy is well known to be a measure of disorder, or of how much uncertainty we have in the whole system.  Its connection to information, its depository and the analysis of what happens in each system has underlying relevance nowadays.

The Gibbs paradox explain the subjective nature of entropy.   It consider a box divided by a wall and we have two equal parts.   Each part has an ideal gas at the same temperature nd pressure.   Without the wall, both parts will mix.
If both gases are different, the entropy of the system will rise.  In the case of equality, both gases are identical and the entropy does not change.  So, the entropy depends on the capacity of the observer to distinguish or not, both gases.  Hence, irreversibility is dependent of what we know about physics \cite{bf}.  Hence, we can conclude that physical objects do not have an intrinsic entropy.

The so-called comoving observers have a system that is dissipationless, i.e., with no gain in the entropy although to the so-called tilted congruence, we have the presence of dissipative fluxes, with entropy producing.

Let us consider a different point of view, by accounting the transitions form the tilted congruence case to the comoving one.   The Landauer principle, 
\cite{landauer} also known as the Brilloin principle \cite{brilloin-varios}, the erasure of one-bit of information within a system demands the dissipation into the environment of a minimal quantity of energy, which if written by
\bee
\label{7}
\delta E\,=\,k_{{}_B} T \ln 2 \;\;,
\eee

\ni where $T$ is the temperature of the environment. The $\ln 2$ term comes directly from the information theory since $e^S=N=2^n$, where $N$ is the dimension of the Hilbert space and $n$ is the total number of Boolean degree of freedom in a region encompassing the BH.   So, the entropy is given by $n\ln 2$.

Namely, Landauer \cite{landauer} demonstrated that reproducing classical information can be carried out reversibly and without wasting energy.   However, when information is erased there is an energy cost greater than $k_{{}_B}\,T\,\ln 2$ per classical bit.   This energy is ultimately converted into thermal energy and raises the final energy of the whole system by $\delta E \ge k_{{}_B}\,T\,\ln 2$.

In other words, erasure means a reset operation recovering the system to a peculiar state, using an external agent, i.e., we can carry out work on it.   However, we have to increase the entropy of another system or the environment.

Hence, the Landauer principle is a reflection of the fact that logical irreversibility implies thermodynamical irreversibility.   It is a particular version of the second law of thermodynamics constructed to include the information theory.   The erasure of data in a system concerns the production of heat and therefore increasing of entropy.   Throughout the erasure of one-bit of information the framework has to generate at least $\delta Q=k_{{}_B}\,T\,\ln 2$ of heat.  Hence, the erasure of one-bit of information raises the classical entropy by $\delta S \ge k_{{}_B}\,\ln 2$.   Landauer principle has been relevant in fathoming the thermodynamics of quantum computation \cite{landauer2,Lloyd,jwzgb} establishing a principle of information erasure for a black hole (BH), that says that ``A BH is one of the most efficient information eraser in systems of a given temperature."   In the information erasing procedure the infalling matter ``saturates the Landauer bound, $\delta E = k_{{}_B}\,T\,\ln 2$" \cite{landauer}.   The result is that this saturating energy $\delta E$ is given to BH, although the system, the intermediate region of information, is in thermal contact with BH that acts as a thermal bath and raise its mass by, as we will see, by
\bee
\label{8}
\delta(Mc^2)\,=\delta E\,=\, k_{{}_B}\,T\,\ln 2\;\;,
\eee

\ni where $T$ is the BH temperature and it multiplies the increased entropy $\delta S =k_{{}_B}\,\ln 2$.  
It is important the notice that the relation in Eq. \eqref{8} does not origin form the first law of BH thermodynamics.   It is understood that the energy $k_{{}_B}\,T\,\ln 2$ is coming from the infalling particle into the system erasing one-bit of information in the system and ultimately to the BH via thermal contact.   Thus, the rise of the BH mass $\delta M$ is connected to the information erasing of the system directly.

Barrow entropy \cite{barrow} has been demonstrated to result in very relevant cosmological phenomena \cite{meus,saridakis}.   It originates from quantum gravitational phenomena on he horizon framework \cite{barrow}, parametrized by the single exponent $\Delta$, where $\Delta=0$ means the standard Bekenstein-Hawking \cite{bekenstein,hawking} entropy and $\Delta=1$ means the maximal fractal deformation.   Consequently, we can realize the at early times these quantum behavior is deeper and so $\Delta$ is closer to one.   However, as time goes by, $\Delta$ tends to its basic value which is zero.   
To sum up, a scenario of Barrow entropy with a variable parameter $\Delta$\footnote{In \cite{blps} the authors suggested a $\Delta$ being dependent of the scale factor.} has a even more reason that the usual energy scale dependence of coupling parameters \cite{blps}.   But to construct a $\Delta=\Delta(M)$ as we make here is completely new in the literature

This paper follows an organization such that in section 2 we used the Barrow model with $\Delta=\Delta(M)$ to compute the entropy, temperature and the addition of the area of the BH horizon where one-bit of information is erased.   In section 3, we obtained the standard expression for this area addition for $\Delta=$constant, i.e., $\Delta=0$ and $\Delta=1$.  In section 4, we analyzed a simple form for the function $\Delta=\Delta(M)$.   Finally, the final discussions, conclusions and perspectives are depicted in section 5.

\section{The Barrow variable parameter}

Barrow \cite{barrow}, based in the geometry of the Covid-$19$ virus, planned a fractal model for BHs where the entropy is given by
\bee
\label{2.1}
S_{{}_{Ba}}\,=\, \left(\frac{k_{{}_B} A}{4 \ell_P^2} \right)^{1+\frac \Delta2}
\eee

\ni Here the deformation provided by quantum gravity is given by the $\Delta$-parameter shown in Eq. \eqref{2.1}, where $\ell_P^2=G$ is the Planck area where $c=\hbar=1$.  From now on we will oscillate showing  relativistic and non-relativistic units in order to indicate pedagogically gravitational, thermodynamical quantum and relativistic features of the specific equation, 
but we will mention where one or the other are been used, in a very simple and direct way, of course.  The value of $\Delta$ is crucial for the understanding what kind of geometry we are dealing with.  For example, for $\Delta=0$ it is direct to see that we have the well known Bekenstein-Hawking entropy in Eq. \eqref{2.1} and for $\Delta=1$ we have he maximal deformation of the BH surface thanks to its fractal structure as suggested by Barrow \cite{barrow}.

To be more specific, the geometry visualized by Barrow was created by an infinite minimizing hierarchy of touching spheres around the Schwarzschild event horizon.  Beginning with a Schwarzschild BH with mass $M$ and radius $R_g=2GM/c^2$, some smaller spheres are attached to its outer area.   Moreover, further smaller spheres are also attached to the firstly mentioned outer spheres and this sequence keeps going to provide a highly intricate fractal structure.   Hence, the boundary will encompass surfaces of hierarchically framework induces a finite volume, but with the finite or infinity area, which in turn provides the modification of the BH entropy given in Eq. \eqref{2.1} \cite{mr}.

This toy model was created basically to show that close the scale where the quantum gravity effects rules \cite{mr} the surface area of a BH are above $4\pi\,R^2_g$ due to the presence of complicated fractal characteristics.   Besides, this happens if we consider any external complexity having a Hausdorff $D\ge 2$ space.   Fundamentally, the $D=2$ geometrical surface behaves like $D > 2$ and tends toward a $D=3$ surface behavior in a limiting scenario with maximal difficulties.  This behavior shows that the $D=2$ surface behaves as if it carries all the information of a $D=3$ volume, the holographic concept.
It must be noticed that this fractal nature does not emerge from any specific quantum gravity computation, but from generalized physical concepts and thus, it is a good proposition as an initial approach \cite{barrow}.

Considering Eq. \eqref{2.1} where $A=4\pi\,R^2_g$ and $R_g=2GM$ and so
\bee
\label{2.2}
A\,=\,16\,\pi\,G^2\,M^2 \;\;.
\eee

Substituting Eq. \eqref{2.2} into Eq. \eqref{2.1} we have that
\bee
\label{2.3}
S_{{}_{Ba}}\,=\, \left(\frac{4\pi k_{{}_B} G^2 M^2}{\ell_P^2} \right)^{1+\frac \Delta2} \;\;.
\eee

\ni So, Eq. \eqref{2.3} can now be written as
\bee
\label{2.4}
S_{{}_{Ba}}\,=\, \Big(4\pi G k_{{}_B} M^2 \Big)^{1+\frac \Delta2} \;\;.
\eee

\ni As we mentioned before, here we will use a $\Delta=\Delta(M)$ form.   From Eq. \eqref{2.4} the first step is to write
\bee
\label{2.5}
\ln S_{{}_{Ba}}\,=\, \Big(1+\frac \Delta2\Big)\,\ln \Big(4\pi G k_{{}_B} M^2\Big)
\eee

\ni and differentiating this equation it yields
\bee
\label{2.6}
\frac{\delta S_{{}_{Ba}}}{\delta M}\,=\,\Big(4\pi G k_{{}_B} M^2 \Big)^{1+\frac \Delta2}\Big[\Big(A_G\,+\,\ln M\Big)\,\frac{\delta \Delta}{\delta M}\,
+\,\frac{(2+\Delta)}{M} \Big] \;\;,
\eee

\ni where $A_G\,=\,\ln\,(4\pi G\,k_{{}_B})/2$

Since we know that
\bee
\label{2.7}
\frac 1T\,=\,\frac{\delta S}{\delta M} \;\;,
\eee

\ni from Eq. \eqref{2.6} we can write that
\bee
\label{2.8}
T\,=\,\frac{1}{\Big(4\pi G k_{{}_B} M^2 \Big)^{1+\frac \Delta2}\Big[\Big(A_G\,+\,\ln M\Big)\,\frac{\delta \Delta}{\delta M}\,
+\,\frac{(2+\Delta)}{M} \Big]} \;\;,
\eee

\ni but, from Landauer's principle, as said above,
\bee
\label{2.9}
\delta E\,=\,k_{{}_B}\,T\,\ln 2 \nonumber
\eee
\bee
\label{2.10}
\Longrightarrow \delta E\,=\,\frac{k_{{}_B}' \ln 2}{M^{2+\Delta}\,\Big[\Big(A_G\,+\,\ln M\Big)\,\frac{\delta \Delta}{\delta M}\,
+\,\frac{(2+\Delta)}{M} \Big]}  \;\;,
\eee

\ni where $$k_{{}_B}' \,=\,\frac{1}{(4\pi G)^{1+\frac \Delta2} k_{{}_B}^{\Delta/2}}$$ and we also know that 
\bee
\label{2.11}
\delta M \,=\, \frac{\delta E}{c^2} \;\;.
\eee

\ni Hence, for $c=1$,
\bee
\label{2.12}
\delta M\,=\,\frac{k_{{}_B}' \ln 2}{\Big(A_G\,+\,\ln M\Big)\,\frac{\delta \Delta}{\delta M}\,
+\,\frac{(2+\Delta)}{M} } \;\;.
\eee

The gravitational radius is given by $R=2GM/c^2$, and consequently the variation of $R$ is $\delta R\,=\,2G(\delta M)/c^2$.   Substituting these values into Eq. \eqref{2.12} we have
\bee
\label{2.13}
\delta R\,=\,\frac{(2G)^{3+\Delta} \ln 2}{(4\pi G)^{1+\frac \Delta2} k_{{}_B}^{\Delta/2} R^{2+\Delta}\Big\{\Big[A_G\,+\,\ln \Big(\frac{R}{2G}\Big)\Big]\,\frac{\delta \Delta}{\delta M}\,
+\,\frac{2G(2+\Delta)}{R} \Big\}} \;\;.
\eee

So, we see clearly that the variation of the radius by an absorption of one-bit of information, in the case of Barrow model, depends directly on the form of the $\Delta$-parameter.   In the next sections we will analyze the results for specific values of $\Delta$, considering $\Delta=$constant, and for a simple form for $\Delta=\Delta(M)$.

\section{The Bekenstein-Hawking and fractal versions of Barrow entropy from a $\Delta$ variable one}

Let us analyze firstly the consequences of adopting a constant value for $\Delta$.   The main values for $\Delta$ are the extremities of the interval $[0,1]$,  we will use these values in Eq. \eqref{2.13} where, for $\Delta=$constant, we have that $\frac{\delta \Delta}{\delta M} =0$,
\bee
\label{2.14}
\delta R_{{}_\Delta=const.}\,=\,\frac{(2G)^{3+\Delta} \ln 2}{(4\pi G)^{1+\frac \Delta2} k_{{}_B}^{\Delta/2} R^{1+\Delta}2G(2+\Delta)}
\eee
\ni and
\bee
\label{2.14-T}
T_{{}_\Delta=const.}\,=\,\frac{1}{(4\pi G k_{{}_B})^{1+\frac \Delta2} (2+\Delta) M^{1+\Delta}} \;\;.
\eee

\ni Thus, we can also write that
\baa
\label{2.15}
4\pi R \,(\delta R) \,=\, \frac{(2G)^{3+\Delta} \ln 2}{2G^2 (4\pi G k_{{}_B})^{\frac \Delta2}(2+\Delta) R^{\Delta}} \nonumber \\
\Longrightarrow \delta A \,=\,\frac{(2G)^{3+\Delta} \ln 2}{G^2 (4\pi G k_{{}_B})^{\frac \Delta2}(2+\Delta) R^{\Delta}} \;\;,
\eaa

\ni where $A$ means the area of the event horizon and the Hawking temperature is given by $$T_{\Delta=0} \,=\, \frac{1}{8\pi G k_{{}_B} M}\;\;,$$ as it is well known in the literature.   

For $\Delta=0$, we have, as mentioned before, the Bekenstein-Hawking entropy and from Eq. \eqref{2.15} we see precisely that the addition of the horizon area due to the erasing of one-bit of information is
\bee
\label{2.16}
\delta A_{\Delta=0}\,=\, 4\,\ell^2_P\,\ln 2 \;\;,
\eee

\ni which means that the area correction given by the erasure of one-bit of information is proportional to the value given by the Landauer principle.  But let us calculate the correction in the case of a fractal perturbation of the geometry.   As mentioned before, $\Delta=1$ represents the maximal fractal deformation of the BH surface and $S_{Ba}\,=\,\Big( \frac{A}{4G}\Big)^{3/2}$.  So, in Eq. \eqref{2.8}, the temperature is 
$$T_{\Delta=1}\,=\,\frac{1}{3\,(4\pi G k_{{}_B})^{3/2}\,M^3}$$ and the correction in the area of the surface is given by
\bee
\label{2.17}
\delta A_{\Delta=1}\,=\,\frac{8\ln 2}{3\sqrt{\pi k_{{}_B}}} \frac{\ell^3_P}{R} \;\;.
\eee

\ni We see that for a large $R$, the addition in the area approximates to zero, although from Eq. \eqref{2.16} we see that $\delta A_{\Delta=0}$ is a constant value independent of $R$.   On the other hand, for $R=\frac{\ell_P}{3\sqrt{\pi k_{{}_B}}}$ we have that both cases are equal, $\delta A_{\Delta=0}= \delta A_{\Delta=1}$.  In relativistic units $R=\frac{\ell_P}{3\sqrt{\pi}}$ and the addition in both zero deformation and maximal deformation are equal, independently of the geometry.

This result can indicate a kind of spin degree of freedom (dof) on all surface components of size $\ell_P^2$.   As we said above, the Landauer theory \cite{landauer} showed that Eq. \eqref{7} is the energy in one-bit of information.  We showed precisely that Eq. \eqref{2.15}, besides being a kind of spin dof of the fractal surface of Barrow BH it can be seen also as the increase of its surface.   Hence, we can say that there is a correspondence between both concepts, the spin dof concept and the Landauer one.   The first came form the holographic principle which means that the Landauer concept can be understood as the first step to the holographic principle.

\section{A simple form for $\Delta=\Delta(M)$}

It is very important to understand that if the environment presents a decreasing of entropy, the system entropy must increase in order to no violate the second law.  It is known in the literature as the generalized second law of thermodynamics of BHs \cite{bekenstein}.

Since the interval of validity of $\Delta$ is $[0,1]$, let us suggest a simple form for $\Delta=\Delta(M)$, namely,
\bee
\label{2.18}
\Delta\,=\,a\,e^{-b M} \;\;,
\eee

\ni where $a$ and $b$ are real positive definite numbers and the interval of validity of $a$  is given by 
\bee
\label{2.18-1}
\{a \in \mathbb{R}_+ / 0\leq a e^{-b M} \leq 1 \} \;\;, 
\eee

\ni which is dependent of the value of $M$.   But the ratio $a/e^{b M}$ must obey the same restriction as the $\Delta$-parameter obeys.   Hence,
\bee
\label{2.19}
\frac{\delta \Delta}{\delta M} \,=\, -\,a b M e^{-b M}
\eee

\ni which shows a decreasing entropy behavior from Eq. \eqref{2.6} and an increasing temperature as shown in Eq. \eqref{2.8}.   The $a=0$ is equivalent to  the $\Delta=$constant case and $0 \leq a \leq 1$.   We can be more specific concerning the value of $a$ if we analyze Eq. \eqref{2.8} using Eq. \eqref{2.18}, so,
\bee
\label{2.20}
T_{a,b}\,=\,\frac{e^{b M}}{\Big( 4\pi G k_{{}_B} \Big)^{1+\frac \Delta2}\,M^{1+\Delta}\Big[2e^{bM}\,+\,a\,-\,a b M \Big(A_G\,+\,\ln M\Big)\Big]}
\eee

\ni and for $\Delta=0$ ($a=0$) in Eq. \eqref{2.18}, hence
\bee
\label{2.21}
T_{{}_{a=0,b}}\,=\,\frac{1}{8\pi G k_{{}_B} M} \;\;,
\eee

\ni the Hawking temperature, as mentioned before.   For $\Delta=1$ which means that $a=1$ and $b=0$ in Eq. \eqref{2.18} we have that
%for a variable $\Delta$ given in Eq. \eqref{2.18}.  For $\Delta=1$,
\bee
\label{2.22}
T_{{}_{a=1,b=0}}\,=\,\frac{1}{3\Big(4\pi G k_{{}_B}\Big)^{3/2}\, M^2} \;\;,
\eee

\ni which shows a temperature of the event horizon for a maximal fractal deformation of the horizon surface.  Consequently, for the increasing in the area of the BH using both Eqs. \eqref{2.18} and \eqref{2.19} we have that
\bee
\label{2.23}
\delta A_{a,b}\,=\,\frac{2(2G)^{3+\Delta}e^{bM}\, \ln 2}
{G(4\pi G k_{{}_B})^{\frac \Delta2} R^{\Delta}\Big\{2G(2e^{bM}+a)\,-\,abR\Big[A_G\,+\,\ln \Big(\frac{R}{2G}\Big)\Big] \Big\}}
\eee
 
\ni and it is direct to see that for $\Delta=0$ ($a=0$), we obtain Eq. \eqref{2.16} and for $\Delta=1$ ($a=1,b=0$), we have Eq. \eqref{2.17}.

Back to the value of $a$ and $b$, let us, for example, analyze Eq. \eqref{2.19}.   Since the temperature is measured in Kevin units, the temperature is positive definite and hence both parameters are connected by the relation
\baa
\label{2.24}
2e^{b M}\,+\,a > a b M \Big(A_G\,+\,\ln M \Big) \nonumber
%\Longrightarrow a < \frac{2e^M}{M^2\Big(A_G\,+\,\ln M \Big) - 1}
\eaa

\ni which means finally that, since $a$ is positive definite,
\bee
\label{2.25}
0 \leq a < \frac{2e^{bM}}{bM\Big(A_G+\ln M\Big)\,-\,1} \;\;,
\eee

\ni since $a$ is also positive definite and $b \in \mathbb{R}_+$.   The more specific condition in Eq. \eqref{2.25} complements the general conditions for $a$ given before in Eq. \eqref{2.18-1}. Notice that in Eq. \eqref{2.19} the condition on $a$ is directly connected to $b$.   In Eq. (28) we have an isolated condition on $a$ and $b \in \mathbb{R}_+$.  The same consideration can be written for $\Delta=1$ since the denominators in Eqs. \eqref{2.19}-\eqref{2.21}, concerning $a$, are all the same.

\section{Conclusions, discussions and perspectives}

As shown in \cite{thooft} the result in Eq. \eqref{2.17} can indicate a kind of spin degree of freedom (dof) on all surface components of size $\ell_P^2$.   As we said above, the Landauer theory \cite{landauer} showed that Eq. \eqref{7} is the energy erasure of one-bit of information when inside a BH.  We showed precisely that Eq. \eqref{2.15}, besides of being a kind of spin dof of the fractal surface of Barrow BH it can be seen also as the increase of the area of its surface.   Hence, we can say that there is a correspondence between both concepts, the spin dof and the Landauer one.   The first came form the holographic principle which means that the Landauer concept can be understood as the first step to the holographic principle.

Moreover, for $\Delta=0$, which recover the Bekenstein-Hawking entropy, as shown in the basics of Barrow model, this is obviously the increasing in the horizon area due one-bit of  information shown in Eq. (19).  On the other hand, for $\Delta=1$, in Eq. (20), it clearly keeps the area dimensions for  $k_{{}_B}=1$, we see that the increasing of Barrow horizon area in its maximal deformation is inversely dependent of $R$ which means that for large $R$ the increasing of the area goes to zero.   It is a completely different situation from the $\Delta=0$ case, which is constant.

%We saw that there is a substantial difference between the addition in the surface of the BH.  Although for Bekenstein-Hawking entropy we have a constant increasing, for the maximal deformation (fractal) it is dependent of the radius of the BH.   Asymptotically we have no increasing in the fractal case.   

To sum up, the increasing in the area due to the erasure of one-bit of information happens in both cases, but for $\Delta=0$ it is constant although for $\Delta=1$ it depends on the radius of BH.

Considering $\Delta=\Delta(M)$, we suggested an exponential form for $\Delta=\Delta(M)$, there are two parameters in this exponential form.   We computed the temperature and the area increasing concerning theses parameters and we established the validity values for both in an independent way. This exponential form. in the case of a decreasing one that we have chosen here, showed also a decreasing behavior of the temperature.   However, given that the temperature is positive definite, it helped us to establish the interval of validity of one of the parameters, as explained just above.   We did not consider Shannon entropy, which is a negative one, as it is the fundamental entropy formulation for the information theory.

\end{document}